\documentstyle[11pt,newpasp,twoside]{article}
\def\edcomment#1{\iffalse\marginpar{\raggedright\sl#1\/}\else\relax\fi}
\reversemarginpar

\begin{document}
\title{Wide-field High-precision CCD Photometry of $\omega$~Centauri and Its RR Lyrae Stars}
 \author{Soo-Chang Rey, Jong-Myung Joo, Young-Jong Sohn, Chang H. Ree, and Young-Wook Lee}
\affil{Center for Space Astrophysics, Yonsei University, Seoul 120-749, Korea}
\author{Alistair Walker}
\affil{National Optical Astronomy Observatory/Cerro Tololo Inter-American Observatory 
(NOAO/CTIO), Casilla 603, La Serena, Chile}

\begin{abstract}
We present wide-field and high-precision $BV$ and $Ca$ \& Str\"omgren $by$ 
photometry of $\omega$~Centauri and its RR Lyrae stars, which represents one of the
most extensive photometric surveys to date for this cluster. The member stars
of $\omega$~Cen are well discriminated from foreground Galactic disk stars from
the different distributions in the $hk$ [=$(Ca-b)-(b-y)$] vs. $b-y$ diagram.
The color-magnitude diagrams show the presence of several distinct red-giant 
branches with a red, metal-rich, sequence clearly separated from other bluer
metal-poor ones. Comparison with our population models suggests the most metal-rich
population is few billion years younger than the most metal-poor population.
From our new $BV$ photometry and $hk$ metallicity measurements for RR Lyrae
stars in $\omega$~Cen, we also confirm that the luminosity of RR Lyrae stars 
depends on evolutionary status as well as metallicity. From the presence of
several distinct populations and the internal age-metallicity relation, we 
conclude $\omega$~Cen was once part of a more massive system that merged with
the Milky Way, as the Sagittarius dwarf galaxy is in the process of doing now.
\end{abstract}

\section{Introduction}

From our CCD photometry with the CTIO 0.9m telescope, we have recently obtained
high-quality homogeneous $BV$ color-magnitude (CM) data for stars in $\omega$~Cen
(Lee et al. 1999). For the first time, we found four distinct red giant-branchs
(RGBs), with a red, metal-rich sequence well separated from other bluer metal-poor
ones. An independent survey by Pancino et al. (2000) later confirmed the reality 
of our initial discovery. The most metal-rich ([Fe/H] $\sim$ -0.5) population in
$\omega$~Cen appears to be few billion years younger than the most metal-poor
([Fe/H] $\sim$ -1.75) population in this system (Lee et al. 1999;
Hughes \& Wallerstein 2000; Hilker \& Richtler 2000).

The multimodal  metallicity distribution  function and  the apparent age-metallicity
relation suggest that the protocluster of $\omega$~Cen was massive enough to
undergo some self-enrichment and several early bursts of star formation. This also
suggests that $\omega$~Cen has evolved within a dwarf galaxy size gas-rich subsystem
until it merged with and was disrupted by our Galaxy few billion years after the formation
of its first  generation metal-poor stars, leaving its core as today's globular cluster
(GC) $\omega$~Cen (see also Majewski et al. 1999; Hughes \& Wallerstein 2000;
Hilker \& Richtler 2000). In this paper, we report our progress in the detailed
analysis of our photometric data.

\section{Observations}

All the observations were made using the CTIO 0.9m telescope and 2K CCD during
six nights in 1996 and three nights in 1997 for $BV$ and $Ca$ \& Str\"omgren $by$
filter systems, respectively. Our observations covered 40 $\times$ 40 arcmin$^{2}$ 
in a 3 $\times$ 3 grid centered on the cluster. In total, 40 - 42 and 2 - 4 frames
were taken in each filter and each field for $BV$ and $Ca$ \& Str\"omgren $by$ 
observations, respectively. Photometry was accomplished using DAOPHOT II and
ALLSTAR (Stetson 1987). We obtained high-quality and homogeneous $BV$ CM data for
more than 130,000 stars (Rey et al. 2001) and light curves for most RR Lyrae stars
(Joo et al. 2001) in $\omega$~Cen.
From the $Ca$ \& Str\"omgren $by$ observations, [Fe/H] abundances for 131 RR Lyrae 
stars were also obtained (see Rey et al. 2000).

\section{Color-Magnitude Diagram}

\begin{figure}      
\begin{center}
\plotfiddle{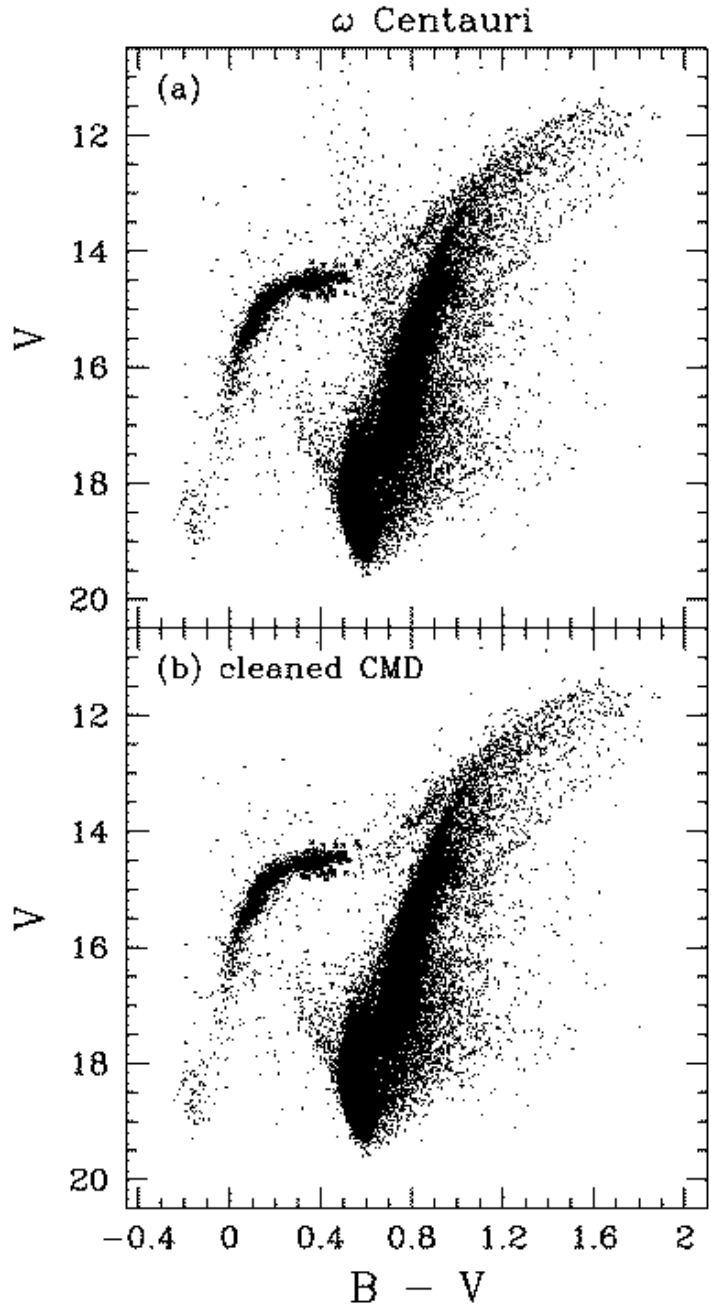}{18cm}{0}{69}{69}{-210}{-12}
\caption{(a) CMD of all stars in our program field with photometric errors
less than 0.05 mag in $B$ and $V$. The RR Lyrae stars are represented by crosses.
(b) Cleaned CMD. A subset of the Galactic field stars have been subtracted using
the $hk$ vs. $b-y$ diagram.}
\end{center}
\end{figure}

\begin{figure} [ht]      
\begin{center}
\plotfiddle{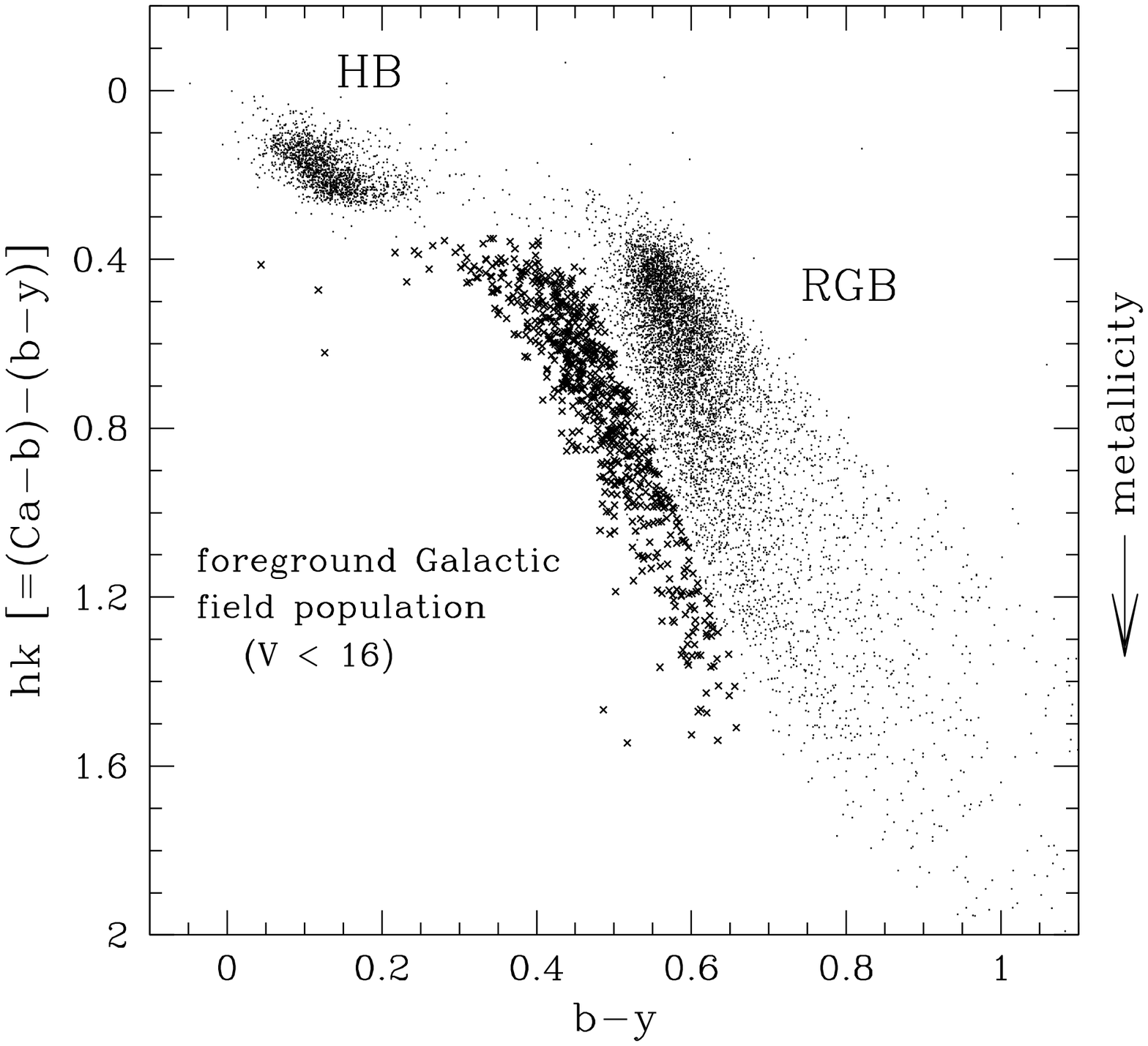}{6.5cm}{0}{43}{43}{-120}{-90}
\caption{$hk$ vs. $b-y$ diagram of stars in the program field with $V$ $<$ 16 mag. 
Galactic field stars (crosses) are clearly discriminated from evolved stars (dots) 
of $\omega$~Cen in this diagram. This is due to the substantial difference between
their mean metallicities.}
\end{center}
\end{figure}

Figure 1a shows a $BV$ CMD for all stars in our program field with photometric
errors less than 0.05 mag in $B$ and $V$. Note the presence of several distinct
RGBs with a red, metal-rich, sequence well separated from other bluer metal-poor ones.
Near $V$ $\sim$ 14.7 and $B-V$ $\sim$ 0.95, a red clump associated with the most
metal-rich population is superimposed by RGB bump, which is shaped as an inclined
sequence by metallicity spread (i.e., fainter as metallicity increases).
The presence of other interesting features on the CMD, such as the blue-tail
phenomenon of the HB and the prominent blue straggler stars, illustrates the diversity
of stellar populations in this cluster.

Because even the metallicity ([Fe/H] $\sim$ -0.5) of the most metal-rich stars in 
$\omega$~Cen are relatively metal-deficient compared to the typical disk field stars, it is  
possible to eliminate foreground field stars through the use of the $hk$ index
(Anthony-Twarog et al. 1991; Twarog \& Anthony-Twarog 1991, 1995).
As shown in Figure 2, at a given $b-y$, more metal-rich stars have larger $hk$ index,
therefore an upper envelope line that cover field stars could be drawn.
Any star that lie below this envelope is tagged as a field star. This implies that 
the use of the $hk$ vs. $b-y$ diagram is ideal for eliminating more metal-rich field 
stars from the $\omega$~Cen, and the most Galactic GC stars as well. Figure 1b shows the
result of the decontamination. The resulting ``cleaned" CMD has allowed us to obtain
more accurate distribution of HB and RGB stars.

\section{Evidence for Multiple Stellar Populations from Red-Giant Branches}

In order to investigate the discrete nature of the RGB, we have plotted in Figure 3
a histogram of the distribution of color difference between  each  RGB star and
the RGB fiducial of the  most  metal-poor component. The RGB stars are selected
in a relatively  narrow magnitude range 12.4 $<$ $V$ $<$ 12.9, so that the field star
and RGB bump contamination is minimized, and which also avoids artificial mixing
in the histogram stemming from metallicity dependence of the RGB  slope. 
The presence of several distinct RGBs is confirmed, although the most metal-rich
component is not as well distinguished as in Fig. 1 because of the
small sample size of such stars in this magnitude range.

\begin{figure} [t]      
\begin{center}
\plotfiddle{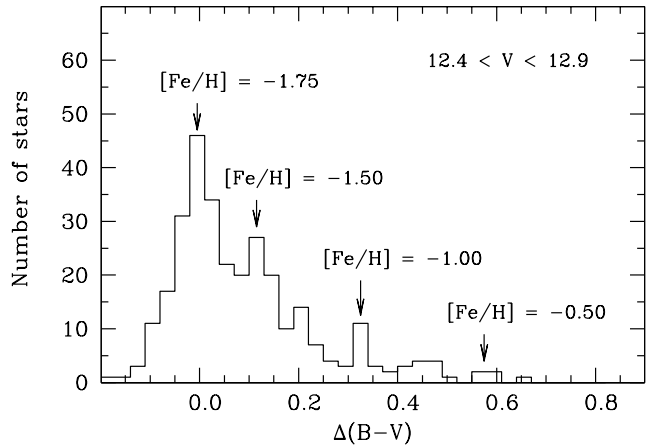}{5cm}{0}{45}{45}{-140}{-125}
\caption{Histogram of the distribution of color difference. Color difference between
each RGB star and RGB fiducial of the most metal-poor component, $\Delta (B-V)$, is
plotted in the range 12.4 $<$ $V$ $<$ 12.9. Note the presence of several distinct
RGBs. The mean [Fe/H] abundances (Zinn \& West scale) for the four main components
are indicated.}
\end{center}
\end{figure}

\section{Age - Metallicity Relation of $\omega$~Centauri}

In our previous analysis using synthetic HB models (Lee et al. 1999), we have made the
relative age estimation from the location of the red clump (i.e., the most metal-rich RHB)
associated with the most metal-rich RGB with respect to the blue HB associated with the
most metal-poor component. 
More detailed analysis (Rey et al. 2001) indicates that this conclusion is somewhat
affected by the presence of RGB bumps associated with several discrete RGBs, 
because of the overlapping of the red clump with RGB bumps at the same location
on the CMD. Therefore, it is important to discriminate the red clump from
metal-poor RGB bump stars for better estimation of relative ages.

\begin{figure}      
\begin{center}
\plotfiddle{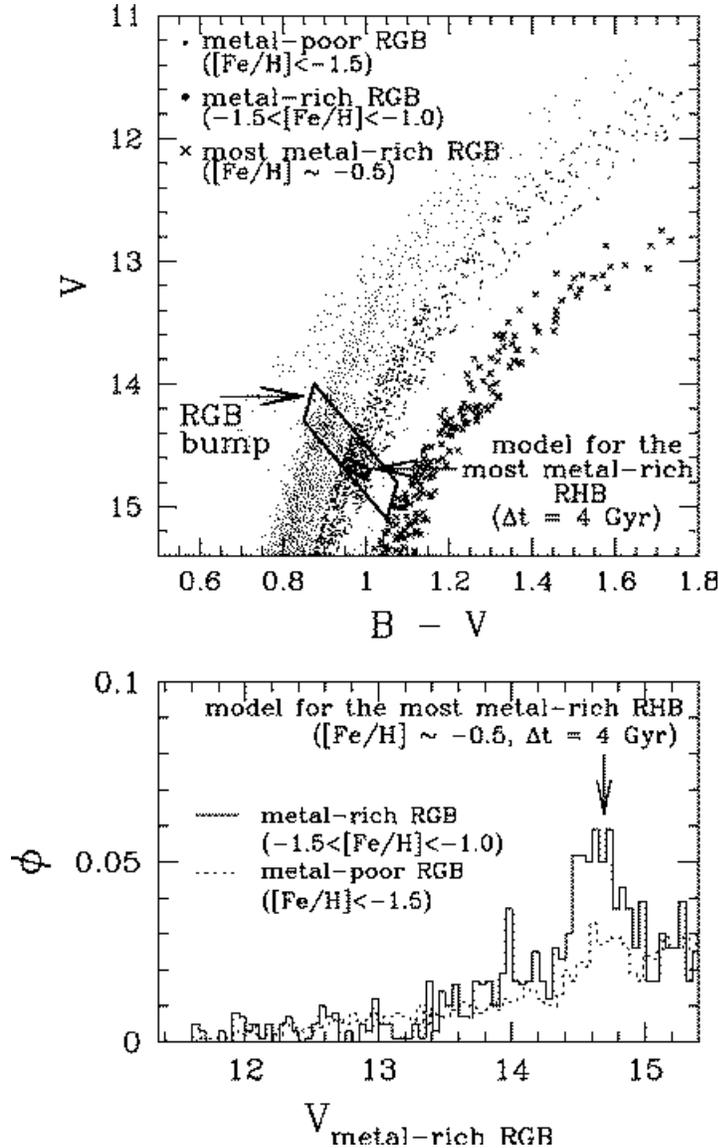}{15cm}{0}{60}{60}{-190}{-10}
\caption{($upper$) CMD that hightlights the RGB and RGB bump region. Three different
RGB sequences with different metallicities are shown. A schematic band of RGB bump
and the predicted location of the red clump from our synthetic HB models
($\Delta t$ $\sim$ 4 Gyr) are indicated. 
($lower$) Normalized LFs for metal-poor (dashed line) and relatively metal-rich
(solid line) RGBs. Peaks of two LFs correspond to the RGB bumps. An arrow indicates
the predicted location of the red clump from our model which is younger
($\Delta t$ $\sim$ 4 Gyr) than the most metal-poor population.}
\end{center}
\end{figure}

Figure 4 shows the CMD that hightlights the RGB and RGB bump region. For the purpose 
of the analysis, RGB is divided into three sequences with different metallicities.
A schematic band of RGB bump is shaped as an inclined sequence by metallicity difference.
A smaller box indicates the predicted location of the red clump from our synthetic HB models
which is younger ($\Delta t$ $\sim$ 4 Gyr) than the most metal-poor population
(see Ree et al., this volume). As shown in the $upper$ $panel$, if the $\Delta t$ $\sim$ 4 Gyr,
the red clump is predicted to be superimposed on the RGB bump of relatively metal-rich
(-1.5 $<$ [Fe/H] $<$ -1.0) population, and therefore we can expect the enhancement of stars 
on the relatively metal-rich RGBs by the red clump stars. In the $lower$ $panel$, we compared
normalized LFs for metal-poor and relatively metal-rich RGBs. Peaks of two LFs correspond
to the RGB bumps. The LF of metal-poor RGB is shifted horizontally in order to coincide
its RGB bump with that of metal-poor RGB. An arrow indicates the predicted location of
the red clump from our model which is younger ($\Delta t$ $\sim$ 4 Gyr) than the most
metal-poor population. Note that the peak of the relatively metal-rich RGB bump is
higher than that of metal-poor one, indicating the red clump stars enhance the feature of
relatively metal-rich RGB bump due to their similar locations. This and the similar analyses
under different assumptions regarding $\Delta t$ suggest that the most metal-rich population
in $\omega$~Cen is some 4 Gyr younger than the most metal-poor population. This conclusion
is in qualitative agreement with the results based on the Str\"omgren photometry of 
main-sequence stars (Hughes \& Wallerstein 2000; Hilker \& Richtler 2000). 
The internal age-metallicity relation is a clear evidence that $\omega$~Cen has enriched
itself over this time scale.

\section{RR Lyrae Stars}

Figure 5a presents the observed M$_{V}$(RR) - [Fe/H] relation for 64 RR$ab$ stars in 
$\omega$~Cen, from our new data for magnitude and metallicity. We adopted a distance 
modulus of $V$ - M$_{V}$ = 14.1 based on the recent evolutionary models of M$_{V}$(RR)
from Demarque et al. (2000). This distance modulus is in good agreement 
with the ones obtained by independent methods (see Kaluzny et al. and Kov\'acs, this volume).
The larger symbols represent stars with smaller observational errors ($\sigma_{[Fe/H]}$ $<$ 0.2 dex)
in [Fe/H]. We have also superposed the model correlation of Lee (1991) with
fixed mass loss and age. The sudden upturn in M$_{V}$(RR) at [Fe/H] $\sim$ -1.5 can be
explained by a series of HB population models (see Fig. 5 of Lee 1993), where one can
see how sensitively the population of the instability strip changes with decreasing [Fe/H].
As [Fe/H] decreases, there is a certain point where the zero age portion of the HB just
crosses the blue edge of the instability strip. Then, only highly evolved stars from the
blue HB can penetrate back into the instability strip, and the mean RR Lyrae luminosity
increases abruptly. This model prediction is clearly visible in our observations.

If the relationship between M$_{V}$(RR) and [Fe/H] is not linear as noticed above,
we can expect a similar correlation between period-shift and [Fe/H]. In order to confirm this,
we obtained the period-shifts of $\omega$ Cen RR$ab$ stars at fixed T$_{eff}$ from the
deviations in the period of each $\omega$~Cen RR$ab$ star from the M3 fiducial line 
in the logP - logT$_{eff}$ plane. In Figure 5b, we present the result with the model
locus by Lee (1993). Our new correlation between period-shift and [Fe/H] shows roughly
the same trend as the M$_{V}$(RR) - [Fe/H] relation, and is in good agreement with
the model locus. All of these results suggest that the luminosity of RR Lyrae stars 
depends on evolutionary status as well as metallicity.

\begin{figure}       
\begin{center}
\plotfiddle{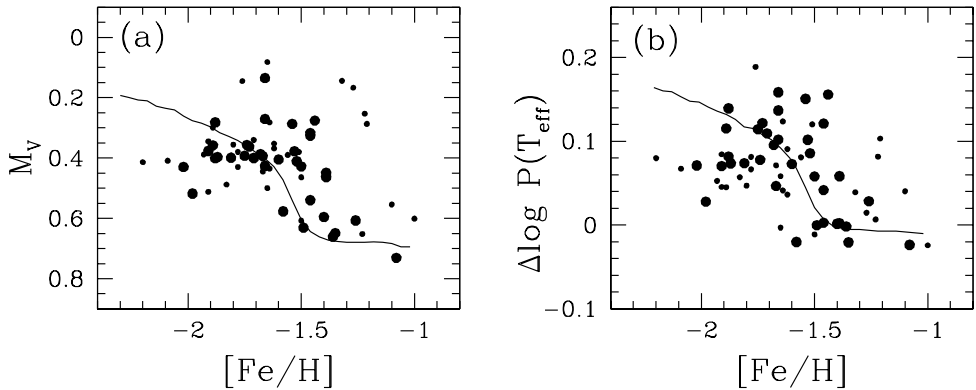}{4.5cm}{0}{65}{65}{-195}{-240}
\end{center}
\caption{(a) M$_{V}$(RR) - [Fe/H] relation of 64 RR$ab$ stars in $\omega$~Cen based on
our $BV$ photometry and metallicity data obtained from our $hk$ method.
(b) Period shift - [Fe/H] relation of 64 RR$ab$ stars in $\omega$~Cen.
Note that our new observations agree well with our model predictions (solid line).}
\end{figure}

\section{Conclusions}
From our study, we draw the following conclusions:

1. The ``cleaned" CMD of $\omega$~Cen shows the presence of several distinct RGBs with
a red, metal-rich, sequence clearly separated from other bluer metal-poor ones.

2. Comparison with our population models suggests the most metal-rich ([Fe/H] $\sim$ -0.5) 
population is few billion years younger ($\Delta t$ $\sim$ 4 Gyr) than the most metal-poor
([Fe/H] $\sim$ -1.75) population. 

3. We present new $BV$ photometry and $hk$ metallicity measurements of RR Lyrae stars
in $\omega$~Cen, which confirm that the luminosity of RR Lyrae stars depends on 
evolutionary status as well as metallicity.

4. From the presence of several distinct populations and the internal age-metallicity
relation, we suggest $\omega$~Cen was once part of a more massive system that merged with
the Milky Way, as the Sagittarius dwarf galaxy is in the process of doing now.

\end{document}